\documentstyle[12pt]{article}
\begin{document}
\large
\begin{center}
{\bf Some Unsettled Questions in the Problem of Neutrino
Oscillations. Remarks About the Majorana Neutrino Oscillations}
\par
\vspace{0.5cm} Beshtoev Kh. M.
\par
\vspace{0.5cm} Joint Inst. for Nucl. Research., Joliot Curie 6,
141980 Dubna,Moscow region, Russia; Inst. of Appl. Math. and
Autom. of the Kabardino-Balkarian Scientific Center of RAC,
Shortanova 89a,
360017 Nalchik, Russia\\
\end{center}

\par
{\bf Abstract}

It is shown  that we cannot put Majorana neutrinos in the standard
Dirac theory without violation of the gauge invariance.
Experiments on $\nu_l \to l (l = e, \mu, \tau)$ transitions do not
confirm this supposition. It means that on neutrino oscillation
experiments, the Majorana neutrino oscillations cannot be observed.\\

\par
{\bf 1. Introduction}\\

\par
At present, it is supposed [1, 2] that the neutrino oscillations
can be connected to Majorana neutrino oscillations. This idea is
based on suggestion that the neutrinos are the Majorana ones i.e.
that the standard weak interactions theory includes only these
neutrinos. Let us come to proof that it leads to contradictions in
the theory and is not confirmed on experiments.\\

\par
{\bf 2. Majorana Neutrino Oscillations}\\

\par
Majorana fermion in Dirac representation has the following form
[1, 3, 4]:
\par
$$
\chi^M = \frac{1}{2} [\Psi(x) + \eta_C\Psi^{C}(x)] , \eqno(1)
$$
$$
\Psi^C(x) \rightarrow \eta_C C \bar \Psi^T(x) ,
$$
\par
\noindent where $\eta_{C}$ is a phase, $C$  is a charge
conjunction, $T$ is a transposition.
\par
From Exp. (1) we see that Majorana fermion $\chi^M$ has two spin
projections $\pm \frac{1}{2}$; and then the Majorana spinor can be
rewritten in the following form:
\par
$$
\chi^M (x) = \left(\begin{array}{c} \chi_{+\frac{1}{2}}(x)\\
\chi_{-\frac{1}{2}}(x) \end{array} \right) . \eqno(2)
$$
The mass Lagrangian of Majorana neutrinos in the case of two
neutrinos $\chi_e, \chi_\mu$ ($-\frac{1}{2}$ components of
Majorana neutrinos and $\bar \chi_{...}$ are  the same Majorana
fermion with the opposite spin projection) in the common case has
the following form:
$$
\begin{array}{c} { \cal L}^{'}_{M} =
 - \frac{1}{2}(\bar \chi_e, \bar \chi_\mu)
\left(\begin{array}{cc} m_{\chi_e} & m_{\chi_e \chi_\mu} \\
m_{\chi_\mu \chi_e} & m_{\chi_\mu} \end{array} \right)
\left(\begin{array}{c} \chi_e \\ \chi_\mu \end{array} \right)
\end{array} .
\eqno(3)
$$
By diagonalizing this mass matrix by standard methods, one obtains
the following expression:
$$
\begin{array}{c}  {\cal L}^{'}_{M} =
 - \frac{1}{2}(\bar \nu_1, \bar \nu_2)
\left(\begin{array}{cc} m_{\nu_1} & 0 \\
0 & m_{\nu_2} \end{array} \right) \left(\begin{array}{c} \nu_1 \\
\nu_2 \end{array} \right) \end{array} , \quad
\begin{array}{cc} \nu_1 = & cos \theta \chi_e - sin \theta \chi_\mu  \\
 \nu_2 = & sin \theta \chi_e + cos \theta \chi_\mu  \end{array} .
 \eqno(4)
$$
These neutrino oscillations are described by standard expressions
(see [1,4,5].
\par
The standard theory of weak interactions is constructed on the
base of local gauge invariance of Dirac fermions. In this case,
Dirac fermions have the following lepton numbers $l_{l,}$ which
are conserved, $ l_{l}, l = e ,\mu , \tau, $ and Dirac
antiparticles have lepton numbers with the opposite sign $ \bar{l}
= - l_{l}$.
\par
Gauge transformation of Majorana fermions can be written in the
following form:
$$
{\chi'}_{+\frac{1}{2}}(x) = exp(-i\beta) \chi_{+\frac{1}{2}}(x) ,
\quad {\chi'}_{-\frac{1}{2}}(x) = exp(+i\beta)
\chi_{-\frac{1}{2}}(x) . \eqno(5)
$$
Then, lepton numbers of Majorana fermions are
\par
$$
l^{M} =\sum_{i} l^{M}_{i} (+1/2) = -\sum_{i} l^{M}_{i}(-1/2) ,
\eqno(6)
$$
\noindent i. e., antiparticle of Majorana fermion is the same
fermion with the opposite spin projection.
\par
Now, we come to discussion of the problem of the place of Majorana
fermion in the standard theory of weak interactions [6].
\par
In order to construct the standard theory of weak interactions
[7], Dirac fermions are used. In this theory, the fermions are
present as doublets. The local current $j^{\mu i}$ of the weak
interaction has the following form:
$$
j^{\mu i} = \bar \Psi_L \tau^i \gamma^\mu \Psi_L  , \eqno(7)
$$
where $\bar \Psi_L, \Psi_L$ are lepton doublets
$$
\left(\begin{array}{c} e\\ \nu_e \end{array}\right)_{i L} \qquad i
= 1-3  , \eqno(8)
$$
where $i$ is flavor number of leptons. If we consider the process
transition of neutrino to corresponding  lepton, we will get the
following reaction:
$$
\nu_l + A(Z) \to l^{-} + A(Z+1) \eqno(9)
$$
The absence of contradictions in this theory with the experimental
data (see [8]) confirms that all fermions are Dirac particles.
\par
Now, if we want  to put the Majorana fermions into the standard
theory, we must take into account that, in the common case, the
gauge charges of the Dirac and Majorana fermions are different
(especially well it is seen in the example of Dirac fermion having
an electrical charge since it cannot have a Majorana charge) . In
this case we cannot just include Majorana fermions in the standard
theory of weak interactions by gauge invariance manner. Then, in
the standard theory the Majorana fermions cannot appear.
\par
In spite of above arguments, if we put Majorana neutrinos in the
standard theory; then on experiments we must see the following
reactions:
$$
\chi_l + A(Z) \to l^{-} + A(Z+1)
$$
with relative probability 1/2 and
$$
\chi_l + A(Z) \to l^{+} + A(Z-1)
$$
with the same relative probability (where $x = e, \mu, \tau$),
since Majorana neutrinos are superpositions of Dirac neutrinos and
antineutrinos. Obviously, all the available experimental data [8]
do not confirm this predictions; therefore we cannot
consider this mechanism as a realistic one for neutrino oscillations.\\

{\bf 3. Conclusion}\\

It was shown  that we cannot put Majorana neutrinos in the
standard Dirac theory without violation of the gauge invariance.
Experiments on $\nu_l \to l (l = e, \mu, \tau)$ transitions do not
confirm this supposition. It means that on neutrino oscillation
experiments
the Majorana neutrino oscillations cannot be observed.\\

\par
{\bf References}\\

\par
\noindent 1. Bilenky S.M., Pontecorvo B.M., Phys. Rep.,
C41(1978)225;
\par
Boehm F., Vogel P., Physics of Massive Neutrinos: Cambridge
\par
Univ. Press, 1987, p.27, p.121;
\par
Bilenky S.M., Petcov S.T., Rev. of Mod.  Phys., 1977, v.59, p.631.
\par
\noindent 2. C. Gonzalez-Garcia 31-st ICHEP, Amsterdam, July 2002.
\par
\noindent 3. Rosen S.P., Lectore Notes on Mass Matrices, LASL
preprint, 1983.
\par
\noindent 4. Beshtoev Kh.M., JINR Communication D2-2001-292,
Dubna, 2001;
\par
\noindent 5. Beshtoev Kh.M., Internet Publ. hep-ph/0103274.
\par
\noindent 6. Beshtoev Kh.M., JINR Commun. E2-92-195, Dubna, 1992.
\par
\noindent 7. Glashow S.L.- Nucl. Phys., 1961, vol.22, p.579 ;
\par
Weinberg S.- Phys.  Rev. Lett., 1967, vol.19, p.1264 ;
\par
Salam A.- Proc. of the 8th Nobel  Symp.,  edited  by N. Svarthholm
\par
(Almgvist and Wiksell,  Stockholm) 1968, p.367.
\par
\noindent 8. Phys. Rev D66, N1, 2002, Review of Particle Physics.

\end{document}